\documentclass{iopart}

\usepackage{ifthen}
\usepackage{ifpdf}

\usepackage{latexsym}
\usepackage{amssymb}
\usepackage{bm}

\ifpdf
\usepackage{graphicx}
\usepackage{epstopdf}
\else
\usepackage{graphicx}
\usepackage{epsfig}
\fi

\newcommand{\eexp}{\mbox{e}^}

\newcommand{\tbox}[1]{\mbox{\tiny #1}}

\newcommand{\amatrix}[1]{\matrix{#1}}

\newcommand{\be}[1]{\begin{eqnarray}\ifthenelse{#1=-1}{\nonumber}{\ifthenelse{#1=0}{}{\label{e#1}}}}
\newcommand{\ee}{\end{eqnarray}}

\newcommand{\hide}[1]{}

\newcommand{\mpg}[2][1.0\hsize]{\begin{minipage}[b]{#1}{#2}\end{minipage}}

\newcommand{\putgraph}[2][width=0.30\hsize]{\includegraphics[#1]{#2}}

\begin{document}

\title[Counting statistics and quantum stirring]
{Counting statistics in multiple path geometries \\ 
and the fluctuations of the integrated current in a quantum stirring device}

\author{Maya Chuchem and Doron Cohen}

\address{Department of Physics, Ben-Gurion University, Beer-Sheva 84105, Israel}


\begin{abstract} 
The amount $Q$ of particles that are transported via 
a path of motion is characterized by its expectation 
value $\langle Q \rangle$ and by its variance $\mbox{Var}(Q)$. 
We analyze what happens if a particle has two optional 
paths available to get from one site to another site, 
and in particular what is $\mbox{Var}(Q)$ for the current which 
is induced in a quantum stirring device. 
It turns out that coherent splitting and 
the stirring effect are intimately related 
and cannot be understood within the framework 
of the prevailing probabilistic theory.
\end{abstract}


\section{Introduction}

The possibility to induce DC currents by periodic (AC) 
modulation of the potential is familiar from the context 
of electronic devices. If an open geometry is concerned, 
it is known as ``quantum pumping" \cite{Thouless,NT84,BPT,Brouwer,AG99,SMCG99},  
while for closed geometry \cite{pmc,pms} we use 
the term ``quantum stirring" \cite{MCWD00,pml}.  
Of recent interest is the possibility to stir 
condensed Bose particles that are confined in 
an optical trap and are described by 
a 3~site Bose-Hubbard Hamiltonian \cite{HKG06,BHK07,SAZ07}.
While in a parallel study \cite{pmb} we explore 
the role of interactions in this stirring process, 
in the present work we would like to explore a new 
aspect of the problem that has to do with {\em counting statistics}.  
For simplicity we neglect the interactions and accordingly 
the problem reduces to ``one particle physics". 

It is well established \cite{blanter,imry,fano1,fano2,fano3,fano4,fano5,fano6,lee,agam} 
that counting statistics in the context of shot noise studies 
provides information on the fluctuations of the occupation 
and on the random probabilistic nature of the quantum-mechanical 
transmission/reflection process. 
In fact these two effects combine together. 
The prototype example is barrier crossing. 
If the average channel occupation is~$f$  
and the transmission probability is~$p$, 
then the emerging number of particles~$Q$ 
is proportional to~$f p$,  
while the variance (per-particle) 
is~${\mbox{Var}(\mathcal{Q}) = (1- f p) f p}$.
Furthermore the results which are derived using classical methods 
(Master equation; Boltzmann-Langevin) are consistent with the 
quantum mechanical calculation (Scattering approach; Green function techniques), 
and the quantum-classical crossover is related to the statistics  
of the transmission coefficients as reflected by 
the Fano factor. \\

{\bf Scope --}
The purpose of this paper is to argue that 
the above common wisdom does not apply in the 
context of quantum stirring. In order to demonstrate 
this point we analyze the prototype 3-site system 
of Fig.~1. We measure the current $\mathcal{I}$ that flows 
through a section in the $c_1$ bond, 
and define the splitting ratio as ${\lambda=c_1/(c_1+c_2)}$.
The integrated current over a time period 
is denoted as  $\mathcal{Q}$. 
If the passage probability from left-to-right is~$p$ 
we do not get for its variance ${\mbox{Var}(\mathcal{Q}) = (1{-}\lambda p) \lambda p}$ 
as implied by the naive probabilistic considerations, 
but rather ${\mbox{Var}(\mathcal{Q}) = \lambda^2 (1{-}p)p}$. 
Then we turn to analyze a full pumping cycle that 
consists of two sequential Landau-Zener adiabatic passages. 
During the first half of the cycle ${\lambda=\lambda_{\circlearrowleft}}$
while during the second half of the cycle ${\lambda=\lambda_{\circlearrowright}}$. 
If ${ \lambda_{\circlearrowright} \ne \lambda_{\circlearrowleft}}$
then it follows that the net integrated current  
is ${\langle \mathcal{Q}\rangle \ne 0}$, 
and we ask what is the variance ${\mbox{Var}(\mathcal{Q})}$. \\

{\bf Observations --}
There are some qualitative observations that are associated 
with our results and we would like to enumerate them 
in advance:    
{\bf (1)}~Coherent splitting of a wavepacket 
does not generate a noisy current in the system;  
{\bf (2)}~The ``fractional charge" of 
a partial wavepacket is determined by the 
splitting ratio, and is physically meaningful. 
{\bf (3)}~The splitting ratio can be greater  
than unity or negative. This has the  
interpretation of having an induced circulating  
current in the system. 
{\bf (4)}~The splitting ratio concept allows  
an easier better understanding of quantum stirring, 
in comparison with the complicated Kubo formalism~\cite{pmc}. 
{\bf (5)}~The fluctuations in $\mathcal{Q}$ 
reflect the non-adiabaticity of the driving cycle.  
{\bf (6)}~The fluctuations of the integrated current 
grow with time as~$t$ and not as~$\sqrt{t}$.
{\bf (7)}~Interference appear differently in 
the calculation of counting statistics when 
compared with the calculation of occupation statistics.

\section{Definitions}

The current operator $\mathcal{I}$ is a conventional 
observable in quantum mechanics. For a single mode  
ring the current through a section $r=r_0$ can be 
expressed using the position and the velocity operators:
\be{3}
\mathcal{I} \ \ = \ \ \hat{v}\delta(\hat{r}-r_0) \Big|_{\tbox{symmetrized}}
\ee
We denote by $Q$ the total number of particles 
that go through the specified section, 
and accordingly define a counting operator:
\be{4}
\mathcal{Q} \ \ = \ \ \int_0^t \mathcal{I}(t')dt'
\ee
Calligraphic letters are used to distinguish 
the~$\mathcal{I}$ and the~$\mathcal{Q}$ operators,  
and $\mathcal{I}(t) \equiv U(t)^{\dag} \mathcal{I} U(t)$, 
where $U(t)$ is the evolution operator.  
Formally we can diagonalize $\mathcal{Q}$,  
find its $Q$~eigenvalues, 
and the associated $|Q\rangle$~eigenstates. 
The {\em initial} state of the system can be 
expanded in the $Q$~representation, 
and accordingly the {\em final} probability distribution 
of~$Q$ at time~$t$ 
is ${\mbox{P}(Q) = |\langle Q|\Psi \rangle|^2 = |\Psi_Q|^2}$. 
This distribution can be characterized by its moments. 
Of particular interest are the expectation value 
and the variance:
\be{6}
\langle \mathcal{Q}\rangle 
& \ =  \ &             
\sum_Q |\Psi_Q|^2 \, Q
\  =  \ 
\langle \Psi|\mathcal{Q}(t)| \Psi \rangle
\\
\mbox{Var}(Q) 
& \ = \ & \langle \mathcal{Q}^2 \rangle - \langle \mathcal{Q}\rangle^2 
\ee
The naive definition of ${\mbox{P}(Q)}$ 
that we have introduced above 
follows Ref.\cite{levitov1} which was later criticized. 
A more careful analysis \cite{levitov2,nazarov} 
of the continuous measurement scheme 
implies that the full counting statistics 
is characterized by the following physically 
measurable quasi-distribution
\be{0}
\mbox{P}_0(Q) = \frac{1}{2\pi} \int 
\left\langle 
\left[ \mathcal{T}\eexp{-i\frac{r}{2}\mathcal{Q}} \right]^{\dag} 
\left[ \mathcal{T}\eexp{+i\frac{r}{2}\mathcal{Q}} \right] 
\right\rangle 
\eexp{-iQr} dr
\ee
The naive mathematical definition  
is obtained if we ignore time ordering:
\be{0}
\mbox{P}(Q) \ = \ \frac{1}{2\pi} \int 
\left\langle 
\eexp{+ir\mathcal{Q}} 
\right\rangle 
\eexp{-iQr} dr
\ = \ \left\langle \delta(Q-\mathcal{Q})\right\rangle
\ee
It is easily verified that $\mbox{P}_0(Q)$  
has the same first and second moments as $\mbox{P}(Q)$.  
Therefore, for the purpose of variance analysis,  
we refer below to the latter (naive) definition. 
For more details about {\em full} counting statistics 
in the context of closed geometries see~\cite{cnz}.

\section{Modeling}

We consider the simple models that are illustrated in Fig.~1. 
The Hamiltonian of the 3~site system is
\be{9}
\mathcal{H} = 
\left( \begin{array}{ccc} 
u & c_1 & c_2 \\ c_1 & 0 & 1 \\ c_2 & 1 & 0  
\end{array}\right) 
\ee 
Without loss of generality we choose the units of time  
such as to have the hopping amplitude per unit time  
between ${|1\rangle}$ and ${|2\rangle}$ equal~1.    
For $c_1=c_2=0$ there are two energy levels $E_{\pm}=\pm1$, 
that correspond to the states ${|1\rangle\pm|2\rangle}$,  
and $E_0=u$ that corresponds to $|0\rangle$.

The prototype scenario that we consider 
later is as follows:
The particle is initially prepared 
in the left site, and the potential~$u$ is 
slowly raised from ${u < 1}$ to ${u > 1}$. 
This induces an adiabatic crossing from 
the left site to the right side.
For completeness we note that in the standard Bose-Hubbard Hamiltonian  
all the hopping amplitudes are negative, and 
accordingly the symmetric level $E_{+}$ is in fact 
the ground state of the double well, 
to which the condensed particles 
are transported from the $E_0=u$ level.

More generally we consider later a quantum stirring scenario.
The transport is measured via the $0{\mapsto}1$ bond,  
and accordingly the current operator is represented by the matrix 
\be{0}
\mathcal{I} =\left( \begin{array}{ccc} 
0 & ic_1 & 0 \\ -ic_1 & 0 & 0 \\ 0 & 0 & 0  
\end{array} \right)
\ee 
Assuming control over the 
couplings $c_1$ and $c_2$ we can design 
a pumping cycle with non-zero net transport ${\langle \mathcal{Q}\rangle \ne 0}$.
See Fig.2 for illustration.

\section{The two level approximation}

Let us assume a driving cycle of period $t_{\tbox{p}}$, 
such that in any moment ${|c_1|,|c_2| \ll 1}$ and ${u \sim 1}$, 
as in Fig.2.
In the strict adiabatic limit the particle 
stays in the same adiabatic level with no leakage 
to the other levels. But if the driving is not strictly 
adiabatic there is some leakage. Below we introduce  
the conditions for neglecting the leakage to~${E_{-}}$. 
Accordingly the dynamics is very well described 
within the framework of a two level approximation: 
\be{10}
\mathcal{H} 
= \left( \begin{array}{cc} u & c  \\ c & 1  \end{array} \right),
\ \ \ \ \ 
\mathcal{I}
= \lambda\left( \begin{array}{cc} 0 & ic  \\ -ic & 0 \end{array} \right)
\ee 
where 
\be{0}
c \ \ \equiv \ \ \frac{1}{\sqrt{2}}(c_1+c_2) \ \ \equiv \ \ \mbox{effective coupling}
\ee
and 
\be{0} 
\lambda \ \ \equiv \ \ \frac{c_1}{c_1+c_2} \ \ \equiv \ \ \mbox{splitting ratio}
\ee
If we had only two sites as in the upper illustration 
of Fig.~1, then $\lambda$ would not emerge.

One can estimate the transition probability from an 
initial adiabatic level $E_n$ to some other adiabatic 
level $E_m$ by writing the Hamiltonian in the adiabatic basis:
if one changes a parameter~$X$ the coupling between 
the adiabatic levels is ${\dot{X}\,[i({\partial \mathcal{H}}/{\partial X})_{nm}/(E_n{-}E_m)]}$. 
Then using leading order perturbation theory with respect to the 
driving rate (${1/t_{\tbox{p}} \propto \dot{X}}$), 
and assuming {\em smooth} temporal variation of the potentials,  
one typically obtains (see e.g. Eq.(\ref{e20}))
\be{12}
P \ \ = \ \ \left|\int_{\tbox{cycle}}  \frac{1}{t_{\tbox{p}}}  f\left( \frac{t}{t_{\tbox{p}}} \right)  \eexp{i\Phi(t)} dt\right|^2
\ \ \sim \ \ \eexp{-\Omega t_p}
\ee 
where ${f()}$ provides the time dependence of the coupling 
between the adiabatic levels, and the dynamical phase $\Phi(t)$
is obtained by integrating over ${E_n(X(t)){-}E_m(X(t))}$.  
The parameter $\Omega$ characterizes the rate of change 
of the dynamics phase. In the absence of avoided crossings~$\Omega$ 
is simply the mean level spacing and we label the result 
as~$P_{\tbox{FGR}}$. This notation implies that we deal 
with an off-resonant Fermi-golden-rule (FGR) transition.
But if we have the avoided crossing of the $E_0$ and the ${E_{+}}$  levels, 
then the predominant contribution to the integral comes from a 
time interval $t_{\tbox{LZ}}=c/\dot{u}$, during which $\dot{\Phi} \sim c$.
Thus for the so-called Landau-Zener transition 
we effectively have in Eq.(\ref{e12}) 
the replacements  $\Omega \mapsto c$ 
and ${t_{\tbox{p}}\mapsto c/\dot{u}}$, 
and we get $P_{\tbox{LZ}} \gg P_{\tbox{FGR}}$. 
The exact result including the correct prefactors is \cite{zener,berry,exact}
\be{21}
P_{\tbox{LZ}} = \exp{\left[ -2\pi\frac{c^2}{\dot u}  \right]}
\ee
which is known as the Landau-Zener transition probability. 
It is important to realize that the first order calculation 
reproduces correctly the singular exponential dependence 
of $P$ on the rate ($\dot{u}$) of the driving.

In this paper we assume an adiabatic stirring process and analyze 
it within the framework of a two level approximation.
This means that two inequalities have to be satisfied: 
\be{0}
P_{\tbox{FGR}} \ll P_{\tbox{LZ}} \ll 1
\ee 
From the above discussion it should be clear that 
for smooth temporal variation of the potential,  
the first inequality is automatically satisfied 
simply because $c \ll 1$ is much smaller compared with 
the mean level spacing, and it can be further improved 
if we care for separation of time scales  ${t_{\tbox{p}} \gg t_{\tbox{LZ}}}$.

\section{Single path crossing}

Consider a 2~site system as in the upper illustration 
of Fig.~1. Initially the particle is prepared in the left site.
Then after some time we measure the probability $p$ 
to find the particle in the right side. In our terminology 
the probability~$p$ characterizes the {\em occupation~statistics} 
at the end of the dynamical scenario. We assume nothing about 
the dynamical scenario, except of being coherent. This means 
that the dynamics is generated by a $2\times2$ Hamiltonian that 
can be possibly time dependent. We ask whether 
the {\em counting statistics} is related to~$p$.   
Within a classical probabilistic point of view 
the answer is very simple: 
one would expect to measure ${Q=1}$ with probability $\mbox{P}(1) = p$, 
and  ${Q=0}$ with probability $\mbox{P}(0)=1{-}p$. 
Accordingly one would expect to have ${\langle \mathcal{Q}^k \rangle = p}$, 
where ${k=0,1,2,...}$. In particular the expectation value  
would be ${\langle \mathcal{Q} \rangle = p}$ and the variance 
would be ${\mbox{Var}(\mathcal{Q})=(1{-}p)p}$.

In the quantum mechanical context this innocent reasoning 
is wrong.  As discussed in Ref.\cite{cnz} the eigenvalues 
of the operator $\mathcal{Q}$ are ${Q_{\pm} = \pm\sqrt{p}}$ 
with probabilities ${p_{\pm} = \frac{1}{2}\left(1\pm\sqrt{p}\right)}$. 
Accordingly the $k$th moment is 
\be{11}
\langle \mathcal{Q}^k \rangle = p_{+}Q_{+}^k +p_{-}Q_{-}^k = p^{\left\lfloor\frac{k+1}{2}\right\rfloor}
\ee 
where $k=0,1,2,3,...$ and $\lfloor...\rfloor$ stands 
for the integer part (i.e. rounded downwards).
Still this result coincides with the corresponding 
classical result for ${k=1,2}$, which we call {\em restricted} quantum-classical correspondence. 
The purpose of the following sections is to explore what happens 
to the relation between {\em counting statistics} and 
{\em occupation~statistics} once multiple path geometries 
are involved, and specifically to evaluate the first two moments 
for a quantum stirring process.

\section{Double path crossing}

Consider a 3~site system as in the lower illustration 
of Fig.~1. Initially the particle is prepared in the left site.
Then we raise slowly the potential from ${u<1}$ to ${u>1}$. 
At the end of the process there is some probability~$p$ to find 
the particle in the right side. Our interest is in the 
counting statistics of the current that flows during 
this process through one bond, say the  ${0\mapsto1}$ bond. 
Since we have a double path geometry we expect a splitting ratio 
that reflects the relative coupling strength $|c_1|^2/|c_2|^2$. 
Using a probabilistic point of view the splitting ratio 
should combine with transition probability (${p\mapsto\lambda p}$), 
leading to 
\be{1}
\mbox{Var}(\mathcal{Q}) \ \ = (1- \lambda p) \lambda p 
\ee   
Thus for a complete transition (${p=1}$) with two equally 
probable paths (${\lambda=1/2}$) we expects to measure 
the maximal variance ${\mbox{Var}(\mathcal{Q})=1/4}$.

In the quantum mechanical context this innocent reasoning 
is wrong. By inspection of Eq.(\ref{e10}) we see that 
the results for the double path crossing can be obtained 
from the results for the single path crossing using~$\lambda$ 
as scaling factor. Namely,  
\be{23}
\langle \mathcal{Q} \rangle \ \ &=& \ \ \lambda p 
\\ \label{e2}
\mbox{Var}(\mathcal{Q}) \ \ &=& \lambda^2 (1-p) p 
\ee
The splitting ratio reflects so-to-say
the relative weight of~$c_1$ in the 
net hopping amplitude~${c_1{+}c_2}$.
For ${c_1=c_2}$ we have ${\lambda=1/2}$, 
which would imply an {\em exact splitting} 
of the wavepacket into two equal pieces. 
In particular we realize that for ${p=100\%}$ transfer 
efficiency the variance in such a case 
would not be~$1/4$ but zero.  
The value~$1/4$ would correspond 
to a probabilistic (rather than exact) splitting 
of the wavepacket. We may say that the correct 
description of the transition from the left 
site to the right side is not 
``the particle has an equal probability to go 
either via the first or via the second path", 
but rather ``the particle goes simultaneously 
via the two paths". Both phrasings are equivalent 
if we have in mind expectation values, 
but only the latter phrasing has the correct connotation 
once counting statistics is considered. 

Naively we expect that a fraction ${0<\lambda<1}$ 
would be transported  via the ${0\mapsto1}$ bond, 
while the complementary fraction ${0<(1{-}\lambda)<1}$   
would be transported via the ${0\mapsto2}$ bond. 
But if $c_1$ and $c_2$ have opposite signs 
then (say) $\lambda$ becomes larger than unity, 
while $(1{-}\lambda)$ is negative. This reflects that 
the driving induces a {\em circulating current} within 
the ring, and illuminates the fallacy of the classical 
peristaltic point of view which we discuss below.

\section{The peristaltic picture}

Typically the driving is periodic, and $Q$ is defined 
as the amount of particles that are transported per 
period. The feasibility to have a non-zero~$\langle \mathcal{Q} \rangle$
(non zero ``DC" current) due to periodic (``AC") driving 
is known in the context of open geometry as ``quantum pumping" \cite{BPT,Brouwer}. 
We use the term ``quantum stirring" \cite{pml,pms} 
in order to describe the analogous effect with regard 
to a closed device. 
During an adiabatic pumping cycle a conventional 
two barrier quantum device takes an electron 
from the left lead and ejects it to the right lead. 
Hence the pumped charge per cycle for a leaky pump 
is naively expected to be $\langle \mathcal{Q} \rangle < 1$. 
This naive result is indeed valid if the pump is operated  
in an open geometry between two reservoirs.

Inspired by the peristaltic picture we assume 
control over the on-site potential~$u$ and the 
coupling constants ($c_1$ and $c_2$) 
which are like ``valves". In the first half of the cycle
${c_2=0}$ and~$u$ is raised across $u\sim1$,  
so as to have an adiabatic passage from the left side 
to the right side via the ${0\mapsto1}$ bond.
In the second half of the cycle ${c_1=0}$ and~$u$ 
is lowered so as to have an adiabatic passage  
from the right side back to the left side via 
the ${0\mapsto2}$ bond.
The net effect is to pump one particle per cycle.

\section{Quantum stirring} 

We can use the results that have been obtained 
for a double path adiabatic passage in order to illuminate 
the fallacy of the peristaltic picture once quantum stirring 
in a closed geometry is considered. If during the first half 
of the cycle ${\lambda=\lambda_{\circlearrowleft}}$, 
and during the second half of the 
cycle ${\lambda=\lambda_{\circlearrowright}}$ 
then in leading approximation (neglecting non-adiabatic effects) we get
\be{25}
\langle \mathcal{Q} \rangle \ \ = \ \ 
\lambda_{\circlearrowleft} - \lambda_{\circlearrowright}
\ \ \ \ \ \ \ \mbox{[per cycle]}
\ee
An optional way to derive this result is to make 
a full 3~level calculation using the Kubo formula \cite{pmc}. 
Here we have bypassed the ``long derivation" 
by making a reduction to an ``adiabatic passage" problem.

It is correct that for a simple minded cycle, 
where either $c_1$ or $c_2$ are zero at 
each stage, we get an agreement with the peristaltic 
picture. But in general this is a fallacy. 
In fact the essence of quantum stirring 
is the circulating current which is induced 
by the driving. 
Contrary to the naive expectation we can get $\mathcal{Q}\gg1$ 
per cycle. This happens if $c_1$ and $c_2$ are roughly 
opposite in sign and hence ${|\lambda|\gg1}$. 
This also can happen if $c_1$ and $c_2$ have the same sign:
if we had considered an adiabatic passage 
at $u\sim-1$ from $|0\rangle$ to the antisymmetric 
state ${|1\rangle{-}|2\rangle}$ 
then it would be like replacing $c_2$ by $-c_2$. 
In general it is better to say that $\mathcal{Q}\gg1$ 
reflects a huge circulating current which is induced
if the pumping cycle encircles a degeneracy \cite{pmc}.

To avoid confusion it should be emphasized that 
we are not talking here about persistent currents 
which are zero order conservative effect,   
but about ``linear response" which is a first order 
effect that might have in general both geometric 
and dissipative aspects. It is important to remember 
that the amount of pumped charge per cycle is well defined 
in the ${\dot{u}\rightarrow0}$ adiabatic limit, 
implying that it does not depend on the actual duration of the cycle.

\section{Fluctuations} 

Having determined $\langle \mathcal{Q} \rangle$ per cycle 
we would like to find out what is the variance ${\mbox{Var}(\mathcal{Q})}$. 
The straightforward calculation procedure is to write the current operator  
in the Heisenberg picture:
\be{0}
\mathcal{I}(t)_{nm}
 \ \ = \ \ 
\langle n | U(t)^{\dag}\mathcal{I}U(t) | m \rangle 
\ee
and then to integrate it over time so as to get 
\be{0}
\mathcal{Q}_{nm} 
\ \ \equiv \ \ 
\left(\amatrix{
+Q_{\parallel} & iQ_{\perp} \cr
-iQ_{\perp}^* & -Q_{\parallel}
}\right)
\ee
The first two moments $\langle \mathcal{Q} \rangle$ and $\langle \mathcal{Q}^2 \rangle$
are obtained from this matrix, leading to the identifications:
\be{0}
\langle \mathcal{Q} \rangle \ \ &=& \ \ Q_{\parallel} \\
\mbox{Var}(\mathcal{Q})  \ \ &=& \ \ |Q_{\perp}|^2 
\ee
For a single path Landau-Zener crossing in a two-site system 
it has been argued in Ref.\cite{cnz} that the first two moments 
should be the same as in the classical calculation. This is 
the {\em restricted} quantum-classical correspondence 
that has been mentioned in Section~5. Consequently one deduces that  
\be{0}
Q_{\parallel} \ \ &=& \ \ 1{-}P_{\tbox{LZ}} 
\\ 
\label{e260}
Q_{\perp} \ \ &=& \ \ \sqrt{(1{-}P_{\tbox{LZ}})P_{\tbox{LZ}}}  \times \mbox{PhaseFactor}
\ee
But we have a multiple path geometry, for which 
restricted quantum-classical correspondence cannot 
be established. Therefore we have to carry out 
the straightforward calculation, which is much 
more complicated. In practice, in order to make 
the calculation manageable, we can rely on the adiabatic 
approximation scheme. Within this framework the evolution operator is 
\be{0}
U(t) \ \approx \ 
\sum_n \Big|n(t)\Big\rangle 
\ \exp\left[-i\int_{t_0}^{t} E_n(t')dt'\right] \ 
\Big\langle n(t_0)\Big|
\ee
and accordingly the time dependent current operator is: 
\be{-1}
\mathcal{I}(t)_{nm}
\ \ &\approx& \ \ 
\langle n(t) |\mathcal{I}| m(t) \rangle 
\times \exp\left[i\int_{t_0}^{t} E_{nm}(t')dt'\right]
\\ 
\label{e26}
\ \ &\equiv& \ \ 
\lambda\left(\amatrix{... & ic\eexp{i\Phi(t)} \cr  -ic\eexp{-i\Phi(t)} & ...} \right)
\ee
If we use the zero order adiabatic states (in $\dot{u}$)
we get for the diagonal terms {\em zero}, because 
the zero order adiabatic states are time-reversal symmetric 
and hence support zero current.
If we use the first order adiabatic states we get 
for the diagonal terms a non-zero result 
with ${Q_{\parallel} = \lambda_{\circlearrowleft}-\lambda_{\circlearrowright}}$.
The details of this ``Kubo" calculation are not included 
here because this result is a-priory expected 
on the basis of the much {\em simpler} analysis of the previous section.

As emphasized in the previous paragraph, 
in the case of multiple path geometry ${Q_{\perp}}$ 
is not related to ${Q_{\parallel}}$, and therefore 
an actual calculation should be carried out.
The good news is that we get from Eq.(\ref{e26}) a non-zero leading order result 
already in the zero order approximation: 
\be{0}
Q_{\perp} \ \ = \ \ 
\int_{-\infty}^{\infty} \lambda c \,\eexp{i\Phi(t)} dt 
\ee
where 
\be{301}
\Phi(t) \ \ = \ \ \int^t \sqrt{(u-1)^2 + (2c)^2} \, dt'
\ee
For a single path Landau-Zener transition in a two site system 
one substitutes $\lambda=1$ and $u=\dot{u}t$. Then it is possible  
to demonstrate, see Ref.\cite{cnz}, that the outcome 
of the integral is $\sim\sqrt{P_{\tbox{LZ}}}$ in agreement with Eq.(\ref{e260}). 
But we have a multiple path geometry for which the result 
is not known a-priori, so we have to stick to the integral and see 
what comes out. If we had only one Landau Zener crossing we 
would get $\sqrt{P_{\tbox{LZ}}}$ multiplied by the splitting ratio $\lambda$.
A full pumping cycle is a sequence of two Landau Zener crossing, 
one at ${t=t_1}$ and the second at ${t=t_2}$.  
Therefore the integral is a sum of two terms, and we get  
\be{251}
\mbox{Var}(\mathcal{Q}) \ \ = \ \ |Q_{\perp}|^2 \ \ = \ \ 
\Big|\lambda_{\circlearrowleft}\eexp{i\varphi_1}  {+} \lambda_{\circlearrowright}\eexp{i\varphi_2}\Big|^2 \, P_{\tbox{LZ}} 
\ee
where $\varphi_1\equiv \Phi(t_1)$ and $\varphi_2\equiv \Phi(t_2)$. 
The result depends on the phase difference ${\varphi = \varphi_2-\varphi_1}$, 
which is determined by the time separation of the two crossings.

For sake of comparison one should realize that the probability~$p$  
to have remnants of the particle in the right side 
is determined by the coherent addition of 
the two Landau-Zener transitions \cite{efrat}:
\be{20}
p \ \ = \ \ \left|\frac{1}{2}\int_{-\infty}^{\infty}   
\frac{\dot{u}/2c}{1+(u/2c)^2}  \eexp{i\Phi(t)} dt\right|^2
\ \ = \ \ \Big|\eexp{i\varphi_1}-\eexp{i\varphi_2}\Big|^2 \, P_{\tbox{LZ}} 
\ee
Here the interference is with opposite sign, 
because $\dot{u}$ in the integrand changes sign.
Thus we see in a tangible way why due to 
interference ${Q_{\perp}}$ is in general 
not trivially related to ${Q_{\parallel}}$.

\section{Long time limit} 

From the above analysis it follows that $\langle \mathcal{Q} \rangle$ 
grows linearly with the number of cycles. An equivalent statement  
is that the eigenvalues $Q_{\pm}$ of the $\mathcal{Q}$ 
operator grow linearly with the number of cycles. 
It follows that for a general preparation also 
the spreading $\sqrt{\mbox{Var}(\mathcal{Q})}$ grows 
linearly with time. Needless to say that the probabilistic 
point of view would predict  ${\propto \sqrt{t}}$
growth of the spreading, on the basis of the central limit theorem.  
Thus we have here again another manifestation of the 
way in which quantum coherent behavior 
differs from its classical stochastic analog. 
If we have good control over the 
preparation we can select the initial state  
to be a Floque eigenstate of the quantum evolution operator. 
For such preparation the linear growth of the spreading 
is avoided, and it oscillates around a residual value.

\section{Fractional ``charge"} 

The derivation of Eq.(\ref{e11}) is based 
on the observation that the eigenvalues 
of the counting operator are $Q=\pm{\sqrt{p}}$. 
Exactly the same fractional value has surfaced 
in the pioneering publication about counting 
statistics \cite{levitov1}, where 
the authors had interpreted it 
as an effective ``fractional charge".
Their observation was immersed in complicated 
diagrammatic calculations involving a many-body system 
of Fermions in an open geometry. In fact their result 
has been largely ignored once realized  \cite{levitov2}
that the naive definition of $\mbox{P}(Q)$ 
is of no physical relevance, 
while $\mbox{P}_0(Q)$ gives no indication 
for ``quantized"  fractional charges.

It is therefore amusing to realize that 
a similar idea might emerge in the present context.
From the above analysis of coherent splitting it follows 
that for ${p=100\%}$ transfer efficiency it is feasible 
to a ``fractional charge" ${\langle \mathcal{Q} \rangle = \lambda}$
with (formally) zero dispersion. 
In fact the measured fraction can be greater 
than unity which we can call ``mega charge" 
or it can be negative. The  observation of 
a ``mega" charge in this context 
simply reflects the presence 
of an induced circulating current, 
which is the essence of the quantum stirring effect.

In any case it should be clear that the notion  
of ``fractional charge" in the context of coherent 
quantum stirring is possibly misleading, 
and we have raised it merely for argumentative purpose.

\section{Summary}

Counting statistics in closed geometries is 
not of classical nature. Even in the simplest problem  
of a coherent transition between 2~sites,   
the full counting statistics comes out different 
compared with the probabilistic expectation. 
Still, the variance comes out the 
same which can be regarded an example 
of restricted quantum-classical correspondence. 
In contrast to that multiple path geometries require 
further reasoning, because there is no simple way 
to deduce the counting statistics.

One observes that the correct description 
of a quantum passage in a double path geometry 
is not  ``the particle has an equal probability 
to go either via the first or via the second path", 
but rather ``the particle goes simultaneously 
via the two paths". Both phrasings are equivalent 
if we have in mind expectation values, 
but only the latter has the correct connotation 
once counting statistics is considered, 
leading to Eq.(\ref{e2}) rather than Eq.(\ref{e1}). 
The coherent splitting of a quantum particle 
is ``exact" rather than probabilistic. 
Furthermore, in a double path adiabatic passage  
one may find that (say) $170\%$ of the particle goes 
via one path, while $-70\%$ goes via the second path.
This reflects the emergence of a circulating current 
which is induced by the driving.

The analysis of adiabatic passage has opened the way 
to figure out what is the counting statistics 
in the quantum stirring problem.
We argue that both the average and the spreading of $\mathcal{Q}$   
grow linearly in time, where the rate of the former 
characterizes the pumping cycle, while the rate of latter   
depends on the quality of the preparation.

The result Eq.(\ref{e251}) that we have obtained 
for $\mbox{Var}(\mathcal{Q})$ 
is exiting because it demonstrates how interference 
gets into the counting statistics calculation
once multiple path geometries are concerned. 
Unlike in the calculation of the transition probability,  
the interference is with a different sign, 
and consequently the {\em counting~statistics}  
becomes unrelated to the {\em occupation~statistics}. 
This should be contrasted with the single path 
crossing problem where non-trivial topology is not involved, 
and consequently the two types of statistics are a-priori related.

We believe that the analysis of counting statistics in closed 
geometries that posses non-trivial topology not only opens 
an interesting direction in the study of quantum stirring, 
but also unmasks some essential physics 
of the counting statistics problem in general.


\clearpage

\ \\ \ \\ 

\noindent
{\bf Acknowledgments:} \\

\noindent
DC thanks Oded Agam (HUJI) for emphasizing 
the importance of studying counting statistics 
in the context of quantum stirring.   
Yuly Nazarov (Delft) is acknowledged 
for highlighting some major observations regarding 
full counting statistics following Refs.\cite{levitov1,levitov2,nazarov}. 
DC thanks Miriam Blaauboer for a most fruitful visit in Delft TU, 
during which the improved version of this work has been worked out. 
We also enjoyed talking with Yaroslav Blanter (Delft), 
Yuval Gefen (Weizmann), and Shmuel Fishman (Technion). 
The research was supported by grants from  
the Deutsch-Israelische Projektkooperation (DIP), 
and the USA-Israel Binational Science Foundation (BSF).

\ \\ \ \\ 



\clearpage

\ \\

\mpg[0.8\hsize]{

\begin{center}
\putgraph[width=0.5\hsize]{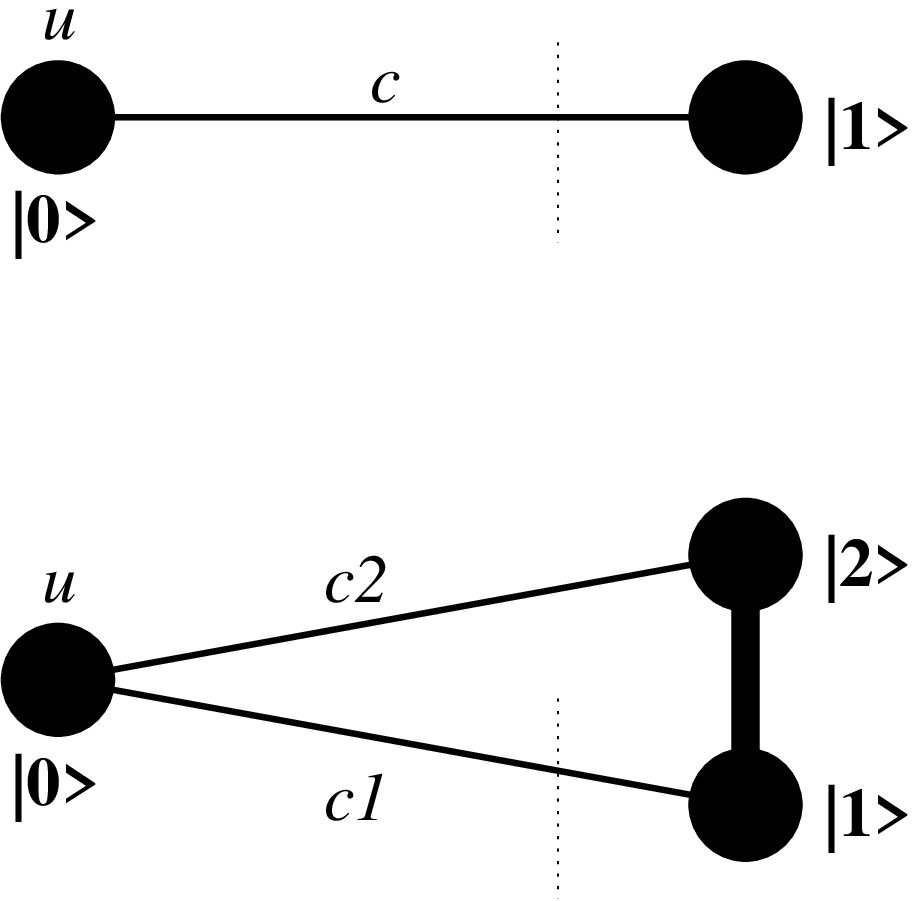}
\end{center}

{\footnotesize 
{\bf Fig.1.}
Toy models that are analyzed in the Letter: 
a particle in a 2~site system (upper illustration), 
and a particle in a 3~site system (lower illustration).
Initially the particle is prepared in the $|0\rangle$ site
where it has a potential energy~$u$.   
The hopping amplitudes per unit time (the~$c$s) 
are also indicated. In the case of a 3~site system, 
the time units are chosen such that the hopping amplitude  
per unit time between $|1\rangle$ and $|2\rangle$ equals unity,  
while the other amplitudes are assumed 
to be small (${|c_1|,|c_2| \ll 1}$). 
The current is measured through the dotted section.} 

}

\ \\ \ \\ \ \\

\mpg[0.8\hsize]{

\begin{center}
\putgraph[width=0.8\hsize]{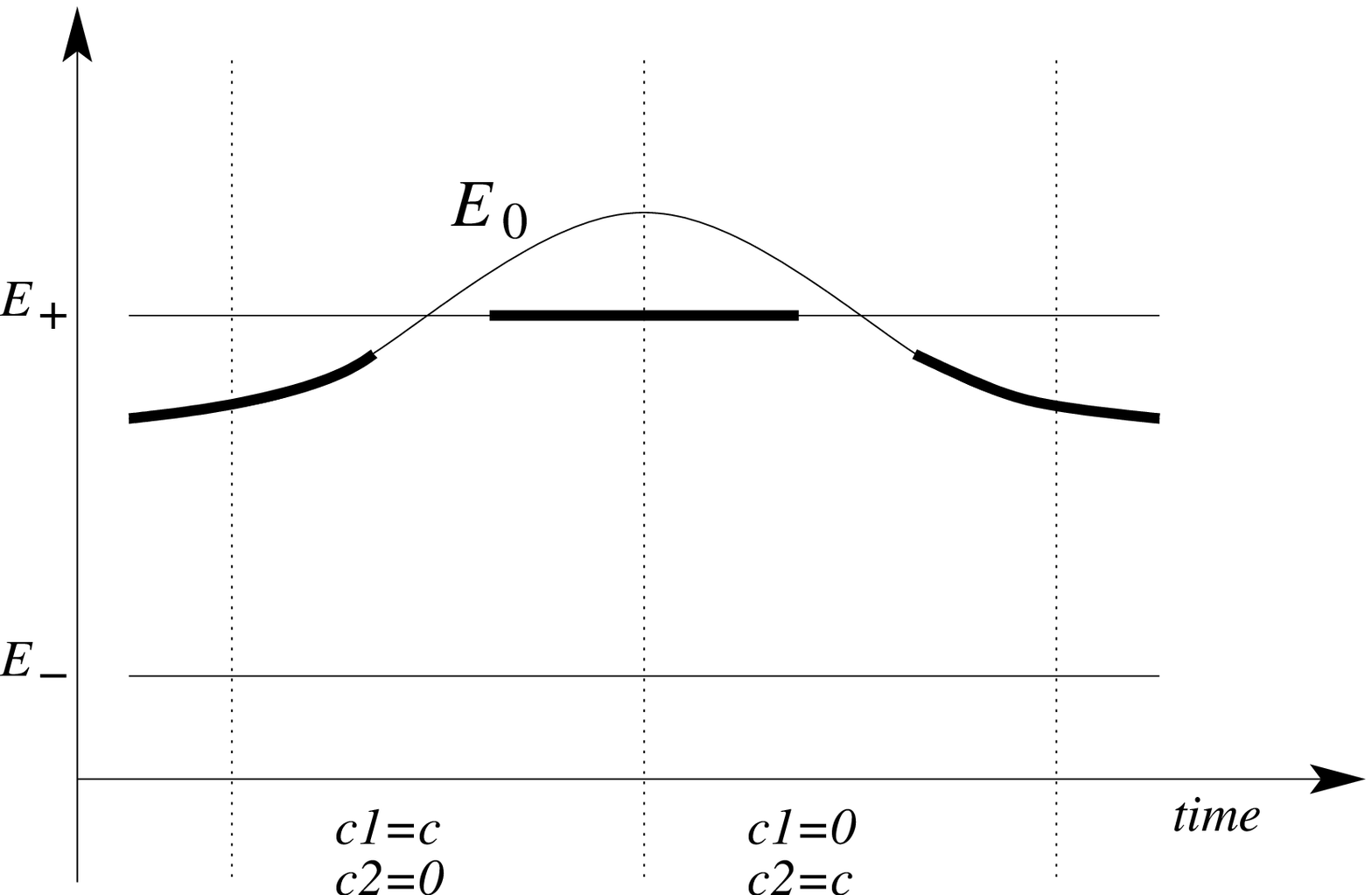}
\end{center}

{\footnotesize 
{\bf Fig.2.}
The adiabatic levels of the 3-site Hamiltonian 
during one period of a pumping cycle. 
In the absence of coupling (${c_1=c_2=0}$) 
the ${E_0=u(t)}$ level intersects  
the symmetric ${E_{+}=1}$ level. 
With non zero coupling these intersections 
become avoided crossings, and the 
particle follows adiabatically the thickened lines. 
For presentation purpose we indicate that 
either $c_1$ or $c_2$ equal zero (``blocked"), 
but in the general analysis we allow any splitting ratio, 
including the possibility ${c_1=c_2}$ of having 
the same amplitude to take either of the two paths.} 

}

\end{document}